\documentclass[preprint,pra,showpacs,showkeys]{revtex4}
\usepackage{amsfonts}
\usepackage{amsmath}
\usepackage{amssymb}
\usepackage{graphicx}
\usepackage{rotating}

\providecommand{\U}[1]{\protect\rule{.1in}{.1in}}

\begin{document}

\preprint{}
\title{Near threshold laser-modified proton emission in nuclear photoeffect }
\author{P\'{e}ter K\'{a}lm\'{a}n$^{1,}$\footnote{%
kalmanpeter3@gmail.com}}
\author{D\'{a}niel Kis$^{2}$}
\author{Tam\'{a}s Keszthelyi$^{1}$}
\affiliation{$^{1}$Budapest University of Technology and Economics, Institute of Physics,
Budafoki \'{u}t 8. F., H-1521 Budapest, Hungary}
\affiliation{$^{2}$Budapest University of Technology and Economics, Institute of Nuclear
Technics, Department of Nuclear Energy, M\H{u}egyetem rkpt. 9., H-1111
Budapest, Hungary}

\begin{abstract}
The change of the probability of proton emission in the nuclear photoeffect due
to an intense coherent (laser) field is discussed near the threshold, where
the hindering effect of the Coulomb field of the remainder nucleus is
essential. The ratio of the laser-assisted and laser-free differential cross
section is deduced and found to be independent of the polarization state of
the $\gamma $ field and the two types of initial nuclear state considered.
The numerical values of this ratio are given at some characteristic
parameters of the intense field.
\end{abstract}

\volumeyear{year}
\volumenumber{number}
\issuenumber{number}
\eid{identifier}
\date[Date text]{date}
\received[Received text]{date}
\revised[Revised text]{date}
\accepted[Accepted text]{date}
\published[Published text]{date}
\startpage{1}
\endpage{}
\maketitle

\section{Introduction}

The development of coherent electromagnetic sources of higher and higher
intensity with increasing photon energy up to the hard x-ray range motivates
the theoretical study of the change of the processes of strongly bound
systems, such as nuclear processes, by these intense fields \cite%
{Ledingham}. In this paper, the change of the nuclear photoeffect due to the
presence of an intense coherent electromagnetic field is studied. This
process is analogous to the laser-assisted x-ray photo effect (x-ray
absorption), a process which was already discussed \cite{KP1} in the late
80's taking into account gauge invariance \cite{Gauge}, \cite{Lamb}. The
laser-assisted nuclear photoeffect (LANP) and the laser-assisted x-ray photo
effect (x-ray absorption) are laser-assisted bound-free transitions. The
difference between them lies in the charged particle (proton or electron,
respectively) which takes part in these processes. Although the LANP was
recently investigated far from the threshold and neglecting the effect of
the Coulomb field of the remainder nucleus \cite{Dadi}, in the case of
the laser-assisted x-ray absorption processes it was found that the most
interesting changes due to the presence of the laser field appear near the
threshold \cite{KP4}, \cite{KP5}.

Thus, applying the results of \cite{KP1}, the LANP is reexamined in a gauge
invariant manner and near the threshold, where the hindering effect of the
Coulomb field of the remainder nucleus is very large so that it must be
taken into account. The effect of the Coulomb field of the remainder nucleus
on the transition rate is approximately taken into account. The
laser-modified differential cross section is compared to the laser-free
differential cross section, and it is shown that their ratio does not depend
on nuclear parameters in the two types of initial nuclear states
investigated and on the state of polarization of the $\gamma $ radiation, but
it has only a laser parameter dependence.

The process investigated can be symbolically written as
\begin{equation}
\omega _{\gamma }\text{ }+\text{ }n\omega _{0}\text{ }+\text{ }_{Z+1}^{A+1}Y%
\text{ }\rightarrow \text{ }_{Z}^{A}X\text{ }+\text{ }_{1}^{1}p,
\end{equation}%
where $_{Z+1}^{A+1}Y$ denotes the target nucleus of mass number $A+1$ and of
charge number $Z+1$. The target nucleus absorbs a $\gamma $ photon
symbolized by $\omega _{\gamma }$, and $n$ laser photons take part
in the process which is symbolized by $n\omega _{0}$. $n<0$ and $n>0$
correspond to $\left\vert n\right\vert $ laser photon emission and
absorption, respectively. As a result, a free proton $_{1}^{1}p$ is emitted
and the remainder nucleus is $_{Z}^{A}X$.

The calculation is made in the radiation $(pA)$ gauge, and in the long
wavelength approximation (LWA) of the electromagnetic fields, the recoil of
\ the remainder nucleus and the initial momentum carried by the laser and $%
\gamma $ fields are neglected. In the case of a circularly polarized
monochromatic wave for the vector potential of a laser field, $\overrightarrow{%
A_{L}}(t)=A_{0}[\cos (\omega _{0}t)\overrightarrow{e}_{1}-\sin (\omega _{0}t)%
\overrightarrow{e}_{2}]$ is used. $\omega _{0}$ is the angular frequency of
the laser. The amplitude of the corresponding electric field $E_{0}=\omega
_{0}A_{0}/c$. The frame of reference is spanned by the unit vectors $%
\overrightarrow{e}_{x}=\overrightarrow{e}_{1}$, $\overrightarrow{e}_{y}=%
\overrightarrow{e}_{2}$ and $\overrightarrow{e}_{z}=\overrightarrow{e}%
_{1}\times \overrightarrow{e}_{2}$. The vector potential describing the
gamma radiation is $\overrightarrow{A}_{\gamma }=\sqrt{2\pi \hbar /\left(
V\omega _{\gamma }\right) }\overrightarrow{\varepsilon }\exp \left( -i\omega
_{\gamma }t\right) $, with $\hbar \omega _{\gamma }$ the energy and $%
\overrightarrow{\varepsilon }$ the unit vector of the state of polarization of
the $\gamma $ photon, and $V$ the volume of normalization.

\section{Gauge invariant S-matrix element of laser-modified proton emission
in nuclear photoeffect}

It is shown in \cite{Gauge} that the electromagnetic transition amplitudes
of a particle (proton) of rest mass $m$ \ and of charge $e$ in the presence
of a laser field are determined by the matrix elements of the operator $-e%
\overrightarrow{r}\cdot \overrightarrow{E}$ with the eigenstates of the
instantaneous energy operator
\begin{equation}
\varepsilon ^{g}=\frac{1}{2m}(\overrightarrow{p}-\frac{e}{c}\overrightarrow{%
A^{g}})^{2}+V(r)  \label{eg}
\end{equation}%
in both ($rE$ and $pA$) gauges. ($e$ is the elementary charge and the
superscript $g$ refers to the gauge.) Accordingly, the gauge-independent
S-matrix element can be written as%
\begin{equation}
S_{fi}=-\frac{i}{\hbar }\int dt\int d^{3}r\psi _{f}^{\ast }\left( -e%
\overrightarrow{r}\cdot \overrightarrow{E}\left( t\right) \right) \psi _{i},
\label{Sfi0}
\end{equation}%
where $\psi _{i}$ and $\psi _{f}$ are the initial and final states of the
proton in the same gauge and $\hbar $ is the reduced Planck constant.

Our calculation is carried out in the radiation $\left( pA\right) $ gauge
because of the choice of the final state of the proton (see below). The
initial state of the proton has the form
\begin{equation}
\psi _{i}=e^{\left( i\frac{e\overrightarrow{r}\cdot \overrightarrow{A}}{%
\hbar c}\right) }\phi _{0}(\overrightarrow{r})e^{-i\frac{E_{b}}{\hbar }t},
\label{pszii}
\end{equation}%
where $\phi _{0}(\overrightarrow{r})$ is a stationary nuclear state of
separation energy $E_{b}$ of the proton. The $\exp \left( i\frac{e%
\overrightarrow{r}\cdot \overrightarrow{A}}{\hbar c}\right) $ factor, where\
$\overrightarrow{A}=\overrightarrow{A}_{L}\left( t\right) +\overrightarrow{A}%
_{\gamma }$, appears because of gauge transformation since $\phi _{0}$ is
the eigenfunction of the instantaneous energy operator,
\begin{equation}
\varepsilon ^{E}=\frac{1}{2m}\overrightarrow{p}^{2}+V_{N}(r)+V_{iC}(r),
\label{eE}
\end{equation}%
in the $rE$ gauge. $V_{N}(r)$ is the nuclear potential and $V_{iC}(r)$ is
the Coulomb potential felt by the proton initially, and the superscript $E$
refers to the $rE$ gauge. The modification of the initial state due to the
laser field is neglected since the direct effect of the intense laser field
on the nucleus has been found to be negligible \cite{Becker} at the laser
parameters discussed. It is also supposed that the initial nucleus does not
have an excited state which is resonant or nearly resonant with the applied $%
\gamma $ radiation.

If similarly to \cite{Dadi} the modification of the final state due to the
strong interaction is neglected, then in the final state and in the $pA$
gauge the instantaneous energy operator $\varepsilon ^{R}$ reads 
\begin{equation}
\varepsilon ^{R}=\frac{1}{2m}(\overrightarrow{p}-\frac{e}{c}\overrightarrow{%
A_{L}}(t))^{2}+V_{C}(r),  \label{eA}
\end{equation}%
where the superscript $R$ refers to the radiation $\left( pA\right) $ gauge
and $V_{C}(r)$ is the Coulomb potential of the remainder nucleus.

An approximated solution of $\left( \ref{eA}\right) $, i.e. an approximated
time dependent state of a particle in the laser plus Coulomb fields, is the
Coulomb-Volkov solution of a proton of wave number vector $\overrightarrow{Q}
$ \cite{Coulomb}, \cite{Rosenberg}:

\begin{equation}
\psi _{\overrightarrow{Q}}\left( \overrightarrow{r},t\right) =V^{-1/2}e^{i%
\overrightarrow{Q}\cdot \overrightarrow{r}}\chi (\overrightarrow{Q}\mathbf{,}%
\overrightarrow{r})\exp \left( -i\widehat{E}t/\hbar \right) f(t).
\label{Volkov}
\end{equation}%
Here $V^{-1/2}e^{i\overrightarrow{Q}\cdot \overrightarrow{r}}\chi (%
\overrightarrow{Q}\mathbf{,}\overrightarrow{r})$ is the Coulomb function,
i.e. the wave function of a free proton in a repulsive Coulomb field of
charge number $Z$, $V$ denotes the volume of normalization, $\overrightarrow{%
r}$ is the relative coordinate of the two particles.
\begin{equation}
\chi (\overrightarrow{Q}\mathbf{,}\overrightarrow{r})=e^{-\pi \eta /2}\Gamma
(1+i\eta )_{1}F_{1}(-i\eta ,1;i[Qr-\overrightarrow{Q}\cdot \overrightarrow{r}%
]),  \label{Hyperg}
\end{equation}%
where
\begin{equation}
\eta \left( Q\right) =Z\alpha _{f}\frac{mc}{\hbar Q},  \label{eta23}
\end{equation}%
is the Sommerfeld parameter, with $\alpha _{f}$ the fine structure constant,
and it is supposed that $m$ is much less than the rest mass of the remainder
nucleus. $_{1}F_{1}$ is the confluent hypergeometric function and $\Gamma $
is the Gamma function \cite{Alder}.

The function
\begin{equation}
f(t)=\exp [i\alpha \sin (\omega _{0}t+\eta _{0})]  \label{ft}
\end{equation}%
where
\begin{equation}
\alpha =\alpha _{\vartheta }\sin \left( \vartheta \right) \text{ \ with \ }%
\alpha _{\vartheta }=\frac{eE_{0}Q}{m\omega _{0}^{2}}.  \label{alpha}
\end{equation}%
Here the polar angles of the wave number vector $\overrightarrow{Q}$ of the
outgoing proton are $\vartheta $ and $\eta _{0}$, i.e. they are the polar
angles of the direction in which the proton is ejected.

In the low-energy range ($QR\ll 1$, where $R$ is the radius of a nucleon)
and for $\left\vert \overrightarrow{r}\right\vert \leq R$ the long
wavelength approximation yields
\begin{equation}
\left\vert \chi (\overrightarrow{Q}\mathbf{,}\overrightarrow{r})\right\vert
_{\overrightarrow{r}=0}=\chi _{C}(Q)=\sqrt{\frac{2\pi \eta \left( Q\right) }{%
\exp \left[ 2\pi \eta \left( Q\right) \right] -1}},  \label{Cb2}
\end{equation}%
which is the square root of the so-called Coulomb factor. (The Coulomb
factor $\left( \chi _{C}^{2}(Q)\right) $ describes well e.g. the Coulomb
correction to the spectrum shape of beta decay \cite{Blatt}.)

For the final state of a proton of wave number vector $\overrightarrow{Q}$,
the LWA of the nonrelativistic Coulomb-Volkov solution $\psi _{%
\overrightarrow{Q}}$ is used.
\begin{equation}
\psi _{\overrightarrow{Q},LWA}\left( \overrightarrow{r},t\right) =\chi
_{C}(Q)V^{-1/2}e^{i\overrightarrow{Q}\cdot \overrightarrow{r}}\exp \left( -i%
\widehat{E}t/\hbar \right) f(t)  \label{pszif}
\end{equation}%
with $\widehat{E}=\hbar ^{2}Q^{2}/(2m)+U_{p}$, that is the energy of the
outgoing proton in the intense field, where $U_{p}=e^{2}E_{0}^{2}/(2m\omega
_{0}^{2})$ is the ponderomotive energy.

Substituting $\left( \ref{pszii}\right) $, $\left( \ref{pszif}\right) $ into
$\left( \ref{Sfi0}\right) $ and using $\overrightarrow{E}=-\frac{1}{c}%
\partial _{t}\overrightarrow{A}$ one can obtain the following form of the
gauge-independent S-matrix element as
\begin{equation}
S_{fi}=-\frac{\chi _{C}(Q)}{\sqrt{V}}\int \exp [i\left( \widehat{E}%
+E_{b}\right) t/\hbar ]f^{\ast }(t)\frac{\partial }{\partial t}G\left[
\overrightarrow{q}\left( t\right) \right] dt,  \label{Sfi}
\end{equation}%
where
\begin{equation}
G\left( \overrightarrow{q}\right) =\int \phi _{0}(\overrightarrow{r})e^{-i%
\overrightarrow{q}\cdot \overrightarrow{r}}d^{3}r  \label{Gq}
\end{equation}%
is the Fourier transform of the initial stationary nuclear state $\phi _{0}(%
\overrightarrow{r})$ of the proton and
\begin{equation}
\overrightarrow{q}\left( t\right) =\overrightarrow{Q}-\frac{e}{\hbar c}%
\overrightarrow{A}.  \label{qt}
\end{equation}%
(Equation $\left( \ref{Sfi}\right) $ can be obtained directly with the aid
of Eq.(27) of \cite{KP1}.)

Using the $\partial _{t}G=\left( \partial _{q}G\right)
\sum_{j=1}^{j=3}\left( \partial _{A_{j}}q\right) \left( \partial
_{t}A_{j}\right) $ identity%
\begin{equation}
\frac{\partial }{\partial t}G=\left( \frac{\partial }{\partial q}G\right)
\frac{e}{\hbar q}\left( \overrightarrow{Q}\cdot \overrightarrow{E}-\frac{e}{%
\hbar c}\overrightarrow{A}\cdot \overrightarrow{E}\right)  \label{dGt}
\end{equation}%
with $\overrightarrow{E}=-\frac{1}{c}\partial _{t}\overrightarrow{A}$, i.e. $%
\overrightarrow{E}=\overrightarrow{E}_{L}\left( t\right) +\overrightarrow{E}%
_{\gamma }$.

The $\overrightarrow{Q}\cdot \overrightarrow{E}_{L}$ term of the last factor
of $\left( \ref{dGt}\right) $ can be neglected if the pure intense field
induced proton stripping process is negligible since this term describes the
process without the gamma photon. Furthermore, the ratio of the amplitudes
of $\overrightarrow{A}_{\gamma }\cdot \overrightarrow{E}_{L}$ and $%
\overrightarrow{A}_{L}\cdot \overrightarrow{E}_{\gamma }$ equals $\omega
_{L}/\omega _{\gamma }\ll 1$. Therefore the $\overrightarrow{Q}\cdot
\overrightarrow{E}-\frac{e}{\hbar c}\overrightarrow{A}\cdot \overrightarrow{E%
}=\overrightarrow{Q}\cdot \overrightarrow{E}_{\gamma }-\frac{e}{\hbar c}~%
\overrightarrow{A}_{L}\cdot \overrightarrow{E}_{\gamma }$ approximation is
justified to use, where $\overrightarrow{E}_{\gamma }=i\sqrt{2\pi \hbar
\omega _{\gamma }/V}\overrightarrow{\varepsilon }\exp \left( -i\omega
_{\gamma }t\right) $. The relative strength of the $\overrightarrow{Q}\cdot
\overrightarrow{E}_{\gamma }$ and $\frac{e}{\hbar c}~\overrightarrow{A}%
_{L}\cdot \overrightarrow{E}_{\gamma }$ terms is characterized by the
parameter $\delta =eA_{0}/\left( \hbar cQ\right) $. In the laser free case $%
Q=\sqrt{2m\Delta }/\hbar $, where $\Delta =\hbar \omega _{\gamma }-E_{b}$ is
the difference of the photon energy and the proton separation energy.
Numerical estimation shows that $\delta \simeq 0.05$ near $\Delta =50$ $keV$
used here and in the case of laser photon energy and intensity values
discussed. Therefore the $\overrightarrow{Q}\cdot \overrightarrow{E}_{\gamma
}$ term is the leading one in the last factor of $\left( \ref{dGt}\right) $.
As to the radiation field dependence of $\partial _{q}G$, the effect of $%
\overrightarrow{A}_{\gamma }$ is negligible in $\overrightarrow{q}(t)$ and
thus $\overrightarrow{q}\left( t\right) =\overrightarrow{Q}-\frac{e}{\hbar c}%
\overrightarrow{A}_{L}$. It was shown above that the amplitude of
oscillation of $\overrightarrow{q}\left( t\right) $ due to the intense field
can be neglected. Therefore $q=Q$ can be used in $\partial _{q}G$, and $%
\left( \ref{dGt2}\right) $ results
\begin{equation}
\frac{\partial }{\partial t}G=\left( \frac{\partial }{\partial q}G\right)
_{q=Q}\frac{e}{\hbar }\frac{\overrightarrow{Q}\cdot \overrightarrow{E}%
_{\gamma }}{q}.  \label{dGt2}
\end{equation}

Using the Jacobi-Anger formula in the Fourier series expansion of $f^{\ast
}(t)$ \cite{Gradstein1} the S-matrix element can be written as%
\begin{eqnarray}
S_{fi} &=&\sum_{n=n_{0}}^{\infty }\frac{2\pi \delta \lbrack \omega _{n}(Q)]i%
}{V}\chi _{C}(Q)\times  \label{Sfi2} \\
&&\times \left( \frac{\partial }{\partial q}G\right) _{q=Q}\frac{e\sqrt{2\pi
\hbar \omega _{\gamma }}}{\hbar }\xi J_{n}(\alpha )e^{-in\eta _{0}},  \notag
\end{eqnarray}%
where $\xi =\overrightarrow{Q}\cdot \overrightarrow{\varepsilon }/Q$, $%
J_{n}(\alpha )$ is a Bessel function of the first kind, and
\begin{equation}
\omega _{n}(Q)=\frac{\hbar Q^{2}}{2m}+\frac{U_{p}+E_{b}}{\hbar }-\omega
_{\gamma }-n\omega _{0}.  \label{OmegaQ}
\end{equation}

The terms, that are small if the $\frac{eE_{0}}{\hbar \omega _{0}\beta _{k}}%
\ll 1$ condition is fulfilled ($\beta _{1}$ and $\beta _{2}$ see below),
were neglected in the calculation.

\section{Gauge invariant differential cross section of laser-modified proton
emission in nuclear photoeffect}

In this paper two cases of initial nuclear energy $-E_{b}$ of the initial
state having a different type of space-dependent part $\phi _{0}\left(
\overrightarrow{r}\right) $ are considered in order to show the general
nature of the effect of the laser on the process. The one case is the $^{8}B$
one-proton halo isotope of separation energy $E_{b}=0.137$ $MeV$ \cite%
{Firestone} and of initial state $\phi _{0}(\overrightarrow{r})=\left( 2\pi
\right) ^{-1}\beta _{1}^{3/2}e^{-\beta _{1}r}/\left( \beta _{1}r\right) $,
with $\beta _{1}=\mu \sqrt{2mE_{b,1}}/\hbar $ and $m$ the rest mass of the
proton $\left( \mu =1.84\text{, }\beta _{1}=1.495\times
10^{12}cm^{-1}\right) $. Although the proton rest mass is more than 12 \% of
the total rest mass of $^{8}B$ and the approximation, the fact that $m$ is much less
than the rest mass of the remainder nucleus, is not very good, but following
\cite{Dadi} we investigate $^{8}B$. In the other case the initial state is
the $\nu s_{1/2}$ shell model state \cite{Greiner} of the form $\phi _{0}(%
\overrightarrow{r})=N_{\nu }e^{-\rho ^{2}/2}F\left( -\nu ,\frac{3}{2},\rho
^{2}\right) $, with $N_{\nu }=\beta _{2}^{3/2}\left[ \Gamma (\nu +\frac{3}{2}%
)/\left( 2\pi \nu !\right) \right] ^{1/2}/\Gamma (\frac{3}{2})$ where $\nu
=0,1,2,...$ is the quantum number of the nuclear shell model and $\Gamma (x)$
denotes the Gamma function. $F\left( -\nu ,\frac{3}{2},\rho ^{2}\right) $ is
the confluent hypergeometric function, $\rho =\beta _{2}r$, $\beta _{2}=%
\sqrt{m\omega _{Sh}/\hbar }$, $m$ is the nucleon rest mass and $\omega _{Sh}$
is the shell model angular frequency ($\hbar \omega _{Sh}=40A^{-1/3}$ $MeV$,
$A$ is the nucleon number \cite{Bohr}, and $\beta _{2}=9.82\times
10^{12}A^{-1/6}$ $cm^{-1}$).

The differential cross section of LANP has the form%
\begin{equation}
\frac{d\sigma }{d\Omega _{q}}=\sum_{n=n_{0}}^{\infty }\frac{d\sigma _{n}}{%
d\Omega _{q}}  \label{Sigma}
\end{equation}%
where $d\Omega _{q}$ is the differential solid angle around the direction of
the outgoing proton. $n_{0}<0$ is the smallest integer fulfilling the $%
\Delta +n\hbar \omega _{0}-U_{p}>0$ condition and $\Delta $ is the same in
both cases of the initial state. (The cases $n<0$ and $n>0$ correspond to $%
\left\vert n\right\vert $ laser photon emission and absorption,
respectively.)

The partial differential cross section
\begin{equation}
\frac{d\sigma _{n}}{d\Omega _{q}}=\sigma _{0,n}\left( Q_{n}\right)
\left\vert \xi \right\vert ^{2}J_{n}^{2}(\alpha _{\vartheta n}\sin \vartheta
)
\end{equation}%
with%
\begin{equation}
\sigma _{0,n}\left( Q_{n}\right) =\alpha _{f}\frac{k_{\gamma }Q_{n}}{2\pi
\lambdabar _{p}}\chi _{C}^{2}\left( Q_{n}\right) \left[ \partial _{Q}G\left(
\overrightarrow{Q}\right) \right] _{Q=Q_{n}}^{2}.
\end{equation}%
Here $Q_{n}=\sqrt{2m\left[ \Delta +n\hbar \omega _{0}-U_{p}\right] }/\hbar $%
, $\xi =\overrightarrow{Q}_{n}\cdot \overrightarrow{\varepsilon }/Q_{n}$, $%
\lambdabar _{p}$ is the reduced Compton wavelength of the proton and $%
k_{\gamma }=\omega _{\gamma }/c$.%
\begin{equation}
Q_{n}=\frac{\varepsilon _{n}}{\lambdabar _{p}}\sqrt{\frac{2\Delta }{mc^{2}}}%
\text{ with }\varepsilon _{n}=\sqrt{1+\frac{n\hbar \omega _{0}-U_{p}}{\Delta
}.}
\end{equation}%
$J_{n}(\alpha _{\vartheta n}\sin \vartheta )$ is a Bessel function of the
first kind, with $\alpha _{\vartheta n}=eE_{0}Q_{n}/\left( m\omega
_{0}^{2}\right) $.

In the variable $\varepsilon _{n}$ the Coulomb factor $\chi _{C}^{2}$\ reads
as%
\begin{equation}
\chi _{C}^{2}(\varepsilon _{n})=\frac{K_{Cb}}{\varepsilon _{n}\left[ \exp %
\left[ \frac{K_{Cb}}{\varepsilon _{n}}\right] -1\right] },  \label{fjk}
\end{equation}%
where $K_{Cb}=2\pi Z\alpha _{f}\sqrt{mc^{2}/\left( 2\Delta \right) }$ with $%
Z $ the charge number of the remainder nucleus. The Coulomb factor causes a
strong hindering of the effect in both the laser assisted and laser free
cases.

\begin{figure}[tbp]
\resizebox{8.5cm}{!}{\includegraphics*{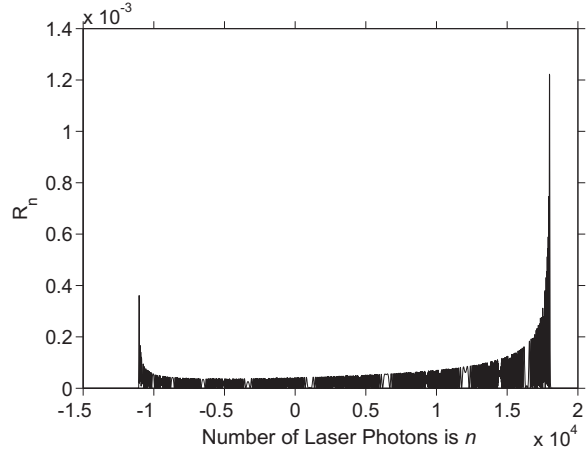}}
\caption{Laser photon number dependence of $R_{n}$ (see $\left( \protect\ref%
{Rnpolj}\right) $), that is the rate of change in one channel due to the
presence of intense field. The cases $n<0$ and $n>0$ correspond to $%
\left\vert n\right\vert $ number laser photon emission and absorption,
respectively. The charge number of the remainder nucleus $Z=4$, $\Delta
=\hbar \protect\omega _{\protect\gamma }-E_{b}$ is the difference of the
gamma photon energy and the proton separation energy, $\protect\vartheta =%
\protect\pi /2$, the laser intensity $I=10^{20}$ $Wcm^{-2}$ and the laser
photon energy $\hbar \protect\omega _{0}=1.65$ $eV$.}
\label{figure1}
\end{figure}

\begin{figure}[tbp]
\resizebox{8.5cm}{!}{\includegraphics*{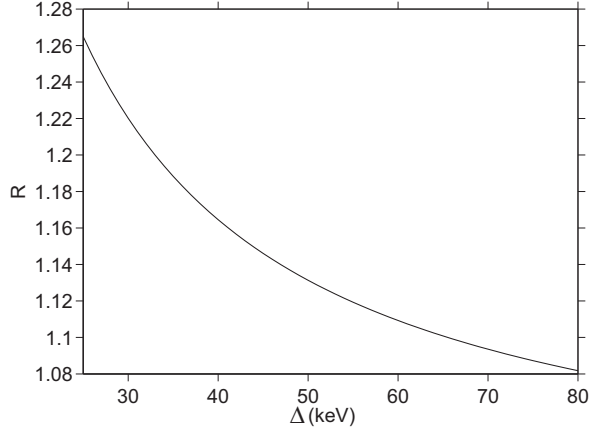}}
\caption{The $\Delta$ dependence of $R$ (see $\left( \protect\ref{R}\right) $%
) at $\protect\vartheta =\protect\pi /2$ in the case of $Z=4$ with $%
I=10^{20} $ $Wcm^{-2}$ and $\hbar \protect\omega _{0}=1.65 $ $eV$. $\Delta
=\hbar \protect\omega _{\protect\gamma }-E_{b}$ is the difference of the
gamma photon energy and the proton separation energy.}
\label{figure2}
\end{figure}

\begin{figure}[tbp]
\resizebox{8.5cm}{!}{\includegraphics*{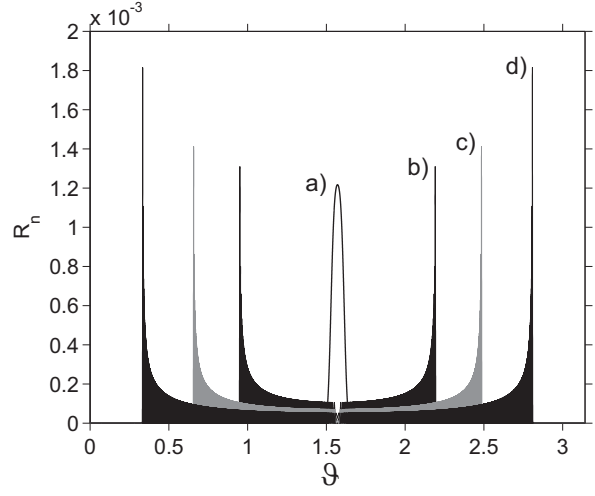}}
\caption{The $\protect\vartheta $ dependence of $R_{n}$ (see $\left( \protect
\ref{Rnpolj}\right) $) with $I=10^{20}$ $Wcm^{-2}$ and $\hbar \protect\omega %
_{0}=1.65$ $eV$ at (a) $n=18000$ (b) $n=14000$ (c) $n=10000$ and (d) $n=5000$%
. The charge number of the remainder nucleus $Z=4$ and $\Delta =50$ $keV$.}
\label{figure3}
\end{figure}

\begin{figure}[tbp]
\resizebox{8.5cm}{!}{\includegraphics*{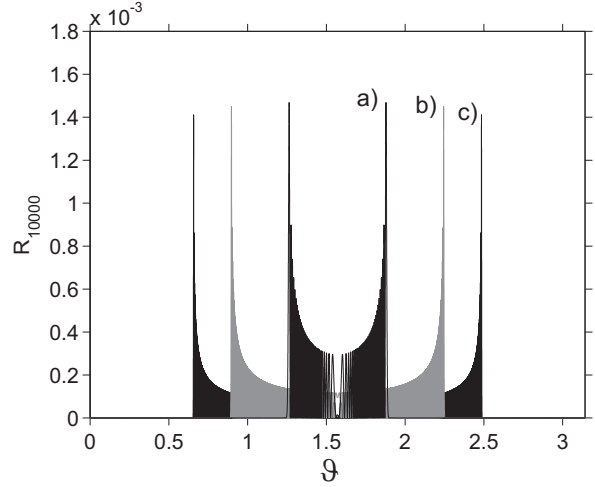}}
\caption{The $\protect\vartheta $ dependence of $R_{n}$ (see $\left( \protect
\ref{Rnpolj}\right) $) with $n=10000$ and $\hbar \protect\omega _{0}=1.65$ $%
eV$ at (a) $I=4\cdot10^{19}$ $Wcm^{-2}$ (b) $I=6\cdot10^{19}$ $Wcm^{-2}$ and
(c) $I=10^{20}$ $Wcm^{-2}$. The charge number of the remainder nucleus $Z=4$
and $\Delta =50$ $keV$.}
\label{figure4}
\end{figure}

Near the threshold $\left( Q_{n}\ll \beta _{1},\beta _{2}\right) $ the $%
\left[ \partial _{Q}G\right] _{Q_{n}}^{2}\varpropto \varepsilon _{n}^{2}$ in
the case of the two types of initial state discussed and
\begin{equation}
\frac{d\sigma _{n}}{d\Omega _{q}}=\sigma _{0}S_{n}\left\vert \xi \right\vert
^{2}.  \label{Sigman}
\end{equation}%
Here $\sigma _{0}$ is constant, it depends on the form of the initial state
and%
\begin{equation}
S_{n}=\frac{\varepsilon _{n}^{2}J_{n}^{2}\left( \alpha _{\vartheta n}\sin
\vartheta \right) }{\exp \left( K_{Cb}/\varepsilon _{n}\right) -1}.
\label{Sn}
\end{equation}

In the laser free case $\left( \alpha _{\vartheta n}=0\right) $ using $%
\varepsilon _{0}\left( U_{p}=0\right) =1$, $J_{0}^{2}\left( 0\right) =1$ and
$J_{n}^{2}\left( 0\right) =0$ at $n\neq 0$, the differential cross section
near above the threshold%
\begin{equation}
\frac{d\sigma ^{th}}{d\Omega _{q}}=\sigma _{0}S_{th}\left\vert \xi
\right\vert ^{2}  \label{SigmaT0}
\end{equation}%
with $S_{th}=\left[ \exp \left( K_{Cb}\right) -1\right] ^{-1}$.

\section{Numerical results}

The ratio $R$ of the laser-assisted and the laser-free differential cross
sections,
\begin{equation}
R=\sum_{n=n_{0}}^{\infty }R_{n},  \label{R}
\end{equation}%
where $R_{n}=S_{n}/S_{th}$ [$\left( \ref{Sigman}\right) $ divided by $\left( %
\ref{SigmaT0}\right) $], equals the ratio of the rates of the corresponding
processes in an elementary solid angle in a given direction of the outgoing
proton. The rate of change in one channel,
\begin{equation}
R_{n}=\frac{\exp \left( K_{Cb}\right) -1}{\exp \left( K_{Cb}/\varepsilon
_{n}\right) -1}\varepsilon _{n}^{2}J_{n}^{2}\left( \alpha _{\vartheta n}\sin
\vartheta \right)   \label{Rnpolj}
\end{equation}%
with
\begin{equation}
\alpha _{\vartheta n}=\varepsilon _{n}\hbar ceE_{0}\left( \hbar \omega
_{0}\right) ^{-2}\sqrt{2\Delta /\left( mc^{2}\right) }  \label{athn}
\end{equation}%
in the variable $\varepsilon _{n}$. So $R_{n}$ and $R$ describe the change
caused by the intense coherent field independently of the state of $\gamma $
polarization and the initial states applied.

In our numerical calculation, the laser photon energy $\hbar \omega _{0}=1.65$
$eV$. First the case in which the outgoing proton moves in the plane of
polarization of the laser beam ($\vartheta =\pi /2$ $\left( \sin \vartheta
=1\right) $) is investigated in the case of $Z=4$. Figure 1 shows the laser
photon number dependence of $R_{n}$ at $\Delta =50$ $keV$ with laser
intensity $I=10^{20}$ $Wcm^{-2}$. The intensity dependence of $R$ has been
investigated with $\Delta =50$ $keV$. $R$ increases linearly from $R=1.00$ \
at $I=10^{18}$ $Wcm^{-2}$ up to $R=1.12$ at $I=10^{20}$ $Wcm^{-2}$. Figure 2
depicts the $\Delta $ dependence of $R$ with $I=10^{20}$ $Wcm^{-2}$. Figure 3
shows the $\vartheta $ dependence of $R_{n}$ in the case $I=10^{20}$ $%
Wcm^{-2}$ at (a) $n=18000$ (b) $n=14000$ (c) $n=10000$ and (d) $n=5000$.
Finally Fig. 4 depicts the $\vartheta $ dependence of $R_{n}$ at $n=10000$
with different laser intensities: (a) $I=4\times 10^{19}$ $Wcm^{-2}$ (b) $%
I=6\times 10^{19}$ $Wcm^{-2}$ (c) $I=10^{20}$ $Wcm^{-2}$ . The numerical
calculation in all the cases discussed above has been repeated at $Z=9$, and
negligible change has been found.

\section{Discussion and summary}

To compare our results and the results of \cite{Dadi} we
investigate our formulas in the $\hbar \omega _{\gamma }\gg E_{b}$ limit. In
this case $\varepsilon _{n}\rightarrow 1$, $\Delta \rightarrow \hbar \omega
_{\gamma }$,
\begin{equation}
\alpha _{\vartheta n}\rightarrow \alpha _{\vartheta }=\hbar ceE_{0}\left(
\hbar \omega _{0}\right) ^{-2}\sqrt{2\hbar \omega _{\gamma }/\left(
mc^{2}\right) }
\end{equation}%
and
\begin{equation}
R_{n}\rightarrow J_{n}^{2}\left( \alpha _{\vartheta }\sin \vartheta \right) .
\end{equation}%
Thus in this limit the $n$ dependence disappears from the argument of the
Bessel function and
\begin{equation}
R\rightarrow \sum_{n=n_{0}}^{\infty }J_{n}^{2}\left( \alpha _{\vartheta
}\sin \vartheta \right) \simeq 1.
\end{equation}

The cross sections obtained in \cite{Dadi} are symmetric in $n$ around $n=0$
(see the figures of \cite{Dadi}) and the total cross section is asserted to
be unaffected by the laser radiation. This corresponds to $R=1$. In
contrast, our result is significantly asymmetric in $n$ ( see Fig. 1.), which
is a consequence of the\ $\varepsilon _{n}$ dependence of $R_{n}$ [see $%
\left( \ref{Rnpolj}\right) $]. The change (increase) of the kinetic energy
of the proton is manifested in the increase of $\varepsilon _{n}$ from $%
\varepsilon _{-11000}=0.761$ up to $\varepsilon _{18000}=1.240$. The
increase of $\varepsilon _{n}$ with increasing $n$ causes an asymmetry in
the $n$ dependence of Fig. 1. The sum of the changes (increments) in the
different channels results in a moderate increment of $R$ ($R<1.28$) as can be
seen in Fig. 2.

Summarizing, one can say that near the threshold, $R$ which measures the
change of the rate of the LANP, has minor $\Delta $ and intensity
dependence, and negligible $Z$ dependence. Furthermore, $R_{n}$, which is the
rate of change in one channel (at a definite laser photon number), has a
significant laser photon number and $\vartheta $ dependences. $R$ and $R_{n}$
are the same in the cases of the two different initial states considered.
Since the numerical results obtained seem to be independent of the initial
nuclear states chosen, it can be expected that $R$ and $R_{n}$ have a minor
dependence from the form of the initial state in general.

Regarding the experimental situation at such a high intensity, it is hard to
distinguish photo-protons from background protons that arise as a result of
interaction with the hot electron plasma created by the intense laser field.
We also have to mention that in an experiment the gamma ray pulse from an
accelerator must be synchronized with an intense (e.g. attosecond) pulse of
a laser system. Moreover, in the case of $^{8}B$, which has a short
half-life of about $770$ $ms$, the $^{8}B$ nuclei must be created in situ in
the laser beam by a nuclear reaction. Fortunately, most of the heavier
nuclei have protons of $\nu s_{1/2}$ shell model state (the other case
investigated) in their stable ground state. The wispy target determined by
the focal spot of the focused intense laser beam, the low repetition rate of
the laser system, and the angular resolution of the proton detector together
make it very challenging to carry out a successful near-threshold, laser-modified
proton emission experiment that could produce significant counting
statistics.

\bigskip


\begin{thebibliography}{99}
\bibitem{Ledingham} K. W. D. Ledingham, P. McKenna, and R. P. Singhal,
Science \textbf{300}, 1107-1111 (2003). [Issn: 0036-
8075; Coden: SCIEAS] [DOI: 10.1126/science.1080552]

\bibitem{KP1} P. K\'{a}lm\'{a}n, Phys. Rev. A \textbf{39}, 2428-2433 (1989). [Issn: 0556-2791; [DOI: 10.1103/PhysRevA.39.2428]

\bibitem{Gauge} R. R. Schlicher, W. Becker, J. Bergou, and M. O. Scully, in%
\textit{\ Quantum Electrodynamics and Quantum Optics}, edited by A. O. Barut
(Plenum, New York, 1984), p. 405.

\bibitem{Lamb} W. E. Lamb, Jr., R. R. Schlicher, and M. O. Scully, Phys.
Rev. A \textbf{36}, 2763-2772 (1987). [Issn:
0556-2791; [DOI: 10.1103/PhysRevA.36.2763]

\bibitem{Dadi} A. Dadi and C. M\"{u}ller, Phys. Rev. C \textbf{85}, 064604
(2012). [Issn: 0556-2813; Coden: PRVCAN]
[DOI: 10.1103/PhysRevC.85.064604]

\bibitem{KP4} P. K\'{a}lm\'{a}n, Phys. Rev. A \textbf{38}, 5458-5460 (1988). [Issn: 0556-2791; [DOI: 10.1103/PhysRevA.38.5458]

\bibitem{KP5} P. K\'{a}lm\'{a}n, Phys. Rev. A \textbf{39}, 3200-3203 (1989). [Issn: 0556-2791; [DOI: 10.1103/PhysRevA.39.3200]

\bibitem{Becker} W. Becker, R. R. Schlicher, and M. O. Scully, Phys. Lett. A
\textbf{106}, 441 (1984). [Issn: 0375-9601;
Coden: PYLAAG] [DOI: 10.1016/0375-9601(84)90989-7]

\bibitem{Coulomb} M. Jain and N. Tzoar, Phys. Rev. A \textbf{18}, 538-545
(1978). [Issn: 0556-2791; [DOI: 10.1103/Phys-
RevA.18.538]

\bibitem{Rosenberg} L. Rosenberg, Phys. Rev. A \textbf{34}, 4567-4574 (1986). [Issn: 0556-2791; [DOI: 10.1103/Phys-
RevA.34.4567]

\bibitem{Alder} K. Alder \textit{et al.}, Rev. Mod. Phys. \textbf{28},
432-542 (1956). [Issn: 0034-6861; Coden: RMPHAT] [DOI:
10.1103/RevModPhys.28.432]

\bibitem{Blatt} J. M. Blatt and V. F. Weisskopf, \textit{Theoretical Nuclear
Physics} (Wiley, New York, 1952), formula (2.10), p. 680.

\bibitem{Gradstein1} I. S. Gradshteyn and I. M. Ryzhik, \textit{Tables of
Integrals, Sums,Series and Products }(Nauka, Moscow, 1971), formula 8.511/3.

\bibitem{Firestone} R. B. Firestone and V. S. Shirley, \textit{Tables of
Isotopes, }8th ed. (Wiley, New York, 1996).

\bibitem{Greiner} J. M. Eisenberg and W. Greiner, \textit{Nuclear
Theory,Vol. 1., Nuclear Models }(North-Holland, Amsterdam, 1970).

\bibitem{Bohr} A. Bohr and B. R. Mottelson, \textit{Nuclear Structure}
(Benjamin, New York, 1969), Vol. 1.
\end{thebibliography}
\end{document}